\documentstyle[twocolumn,floats,aps,psfig]{revtex}
\begin{document}
\draft

%\twocolumn[
\title{Beyond Mean Field \\
Confrontation of Different Models with High Transverse Momentum 
Proton Spectra}
%%%%%%%%%%%%%%%%%%%%%%%%%%%%%%%%%%%%%%%%%%

%%%%%%%%%%%%%%%%%%%%%%%%%%%%%%%%%%%%%%%%%%%%%%%%%%%%%%%%
\author{M. Germain, Ch. Hartnack, J.L. Laville
 and J. Aichelin }
%%%%%%%%%%%%%%%%%%%%%%%%%%%%%%%%%%%%%%%%%%%%%%%%%%%%%
%%%%%%%%%%%%%%%%%%%%%%%%%%%%%%%%%%%%%%%%%%%%%%%%%%%%%%%%%%%
\address{SUBATECH \\
Laboratoire de Physique Subatomique et des Technologies Associ\'ees \\
UMR Universit\'e de Nantes, IN2P3/CNRS, Ecole des Mines de Nantes\\
4, rue Alfred Kastler
F-44070 Nantes Cedex 03, France.}

\author{M. Belkacem}

%%%%%%%%%%%%%%%%%%%%%%%%%%%%%%%%%%%%%%%%%%%%%%%%%%%%%%%%%%%%

\address{Institut f\"ur Theoretische Physik, Johann Wolfgang
Goethe-Universit\"at \\
Postfach 11 19 32, D-60054 Frankfurt, Germany}

%%%%%%%%%%%%%%%%%%%%%%%%%%%%%%%%%%%%%%%%%%%%%%%%%%%%%%%%%%%
\author{E. Suraud
\thanks{Membre de l'Institut Universitaire de France}
}
\address{
Laboratoire de Physique Quantique, Universit\'e Paul Sabatier, 118 route
de Narbonne, F-31062 Toulouse Cedex, France}

%%%%%%%%%%%%%%%%%%%%%%%%%%%%%%%%%%%%%%%%%%%%%%%%%%%%%%%%%%%

%\widetext
%%%%%%%% I know that this is not the best style, but quick and dirty to
%%%%%%%  avoid these shxxx problems with twocolumns and the lost floats..
\author{\begin{quote}
\begin{abstract}
Several models have been proposed to simulate heavy ion reactions beyond
the mean field level. The lack of data in phase space regions which may be
sensitive to  different treatments of fluctuations made it difficult to
judge these approaches.
The recently published high energy proton spectra, measured in the reaction
94 AMeV Ar + Ta, allow for the first time for a comparison of the models with
data. We find that these spectra are reproduced by
Quantum Molecular Dynamics (QMD) and Boltzmann \"Uhling Uhlenbeck (BUU)
calculations. Models like Boltzmann Langevin (BL) in which additional
fluctuations in momentum space are introduced overpredict the proton 
yield at very high energies. The BL approach has been successfully used to
describe the recently measured very subthreshold kaon production assuming
that the fluctuations provide the necessary energy to overcome the
threshold in two body collisions.
Our new findings suggest that the very subthreshold kaon production
cannot be due to two body scattering and thus remains a open problem.
\end{abstract}
%
%\pacs{}
\end{quote}} 
%\narrowtext
% ]
\maketitle
%%%%%%%%%%%%%%%%%%%%%%%%%%%% Introduction%%%%%%%%%%%%%%%%%%%%%%%%
\section{Introduction}
The emission of particles of extremely high energies as well as the
production of particles at beam energies per nucleon far below the threshold
in NN collisions are topics of special interest in the field of heavy ion
collisions at intermediate energies. These processes require a strong
collectivity of the system or at least a high degree of correlated
multiple interactions. An analysis of those processes therefore allows
for a study of fluctuations and correlations in the nuclear reactions.

On the theoretical side this is also a very interesting subject because
it allows to study the predictions of theoretical simulations beyond
mean field level. Mean field calculations have been advanced a long time
ago. Time Dependent Hartree Fock calculations allowed to study the
kinematics of heavy ion reactions at very low beam energies where
two body collisions are negligible\cite{tdhf}. Later, BUU
models, like the Boltzmann
\"Uhling Uhlenbeck model (BUU) \cite{ai85a,cas90}, the Vlasov
\"Uhling Uhlenbeck model (VUU) \cite{st86}, the Boltzmann Nordheim Vlasov 
 (BNV) \cite{bonas}  and the Landau Vlasov model
(LV) \cite{greg87} combined
mean field calculations with a collision term. They succeeded to describe
several observables at beam energies between 20 AMeV and 1 AGeV.

To go beyond the mean field approach is rather challenging. We are only
aware of two approaches which resulted in quantitative predictions of
observables. The so-called Boltzmann-Langevin (BL) approach \cite{su,gr,ra,re}
relies on a stochastic extension of extended (BUU-like) mean-field
theories. Although reasonably well founded from the formal point of view
(essentially as well as its progenitor BUU), the
application of BL to heavy-ion collisions remains extremely
painful, and only approximate methods are presently available 
in realistic cases \cite{su}.
Still, applications of BL have already been proposed for intermediate
mass fragment or sub threshold particle production \cite{su}.
In the case of kaons, a BL-inspired model has even allowed to provide the 
order
of magnitude of kaon production cross sections around 100 AMeV \cite {be}.
The other approach, the so-called Quantum Molecular Dynamics model (QMD)
simulates the quantal many-body problem in an approximate (semi classical)
way. In this model the kaon production below
100 AMeV is zero and is hence in contradiction with experiment.

Recently, first experimental results on the production of high energy protons
in Ar+Ta at 94 AMeV \cite{ge} have been reported. These protons,
having an energy of several times the beam energy, are presumably produced in
collisions with a large $\sqrt{s}$ and hence, possibly,  in the same type of
collisions as the kaons which require a still larger $\sqrt{s}$ value.
These data allow now for the first time to study the predictions of
the two models in a phase space region where the different approaches
to the fluctuations may become relevant. To report about
this comparison is the purpose of this letter.

\section{QMD and BL}

The QMD model is a n-body theory which is based on a variational principle.
It allows to reduce the time evolution of a
n - body test wave function to the time evolution of its parameters.
In QMD the
test wave function is a direct product of n Gaussians. The 6n
parameters are the centroids in coordinate and momentum space of the 
Gaussian wave functions. The time evolution of these centroids are given
by Euler Lagrange equations derived by the variation of the
Lagrangian. They  have the same structure as the classical Hamilton 
equations. The time evolution of the centroids in momentum space is
governed
by almost the same potential as employed in BUU or BL approaches.
The initial condition is chosen in a way to reproduce the coordinate
and momentum space distribution of a cold nucleus. The potential interaction
between the nucleons is supplemented by an elastic and an inelastic
two body scattering. For details we refer
to references \cite{ai,ha}. The calculations reported here were performed with
IQMD, a QMD version which includes isospin explicitly\cite{iqmd}. As far as the
single particle spectra are concerned we do not expect a significant difference
between the different QMD flavours as long as the NN cross section is not 
changed.

The Boltzmann Langevin (BL) theory is a one body mean field theory improved
by incorporating
dynamical fluctuations. A correlation function in momentum space,
which can entirely be derived
from one body properties and which fulfills the fluctuation dissipation
theorem, is employed in each time step of the simulation to spread the
trajectory in momentum space around its original value.
Out of the distribution of new trajectories
one is chosen by a Monte Carlo procedure and propagated to the next time step
where the procedure repeats itself. {\bf  There are different ways to realize 
this procedure in numerical simulations \cite{chapelle}. Only one of these
propositions has been developed to a simulation program which allows 
quantitative comparisons with experiments\cite{suraud}}. In this realization 
one calculates
at each time step the actual (fluctuating) value of the quadrupole
(octupole) moment of the momentum distribution of nucleons being close
together in coordinate space. According to this fluctuating value
of the multipole moment and respecting basic
energy and momentum conservation laws, one assigns to each of the
particles a new momentum with which it
is propagated in the next time step. This procedure allows in rare cases
that a large part of the momentum of many fellow nucleons is
transferred to a single nucleon.
Effectively this corresponds to a many body scattering in which
one of the scattering partners may carry a large fraction of the total (local)
available momentum. This effective many body scattering is not present in
the QMD model. In this respect comparing our BL and QMD simulations allows to
disentangle the relevance of collective effects on a given observable.

It should be noted that the above described BL simulations dubbed {\bf  $BL_{PM}$ } (
BL realized with the so-called "projection method") are by construction devised to properly 
account for quadrupole dominated fluctuations. In this respect these 
simulations do emphasize the collective part (in p-space, not in r-space,
where  they are local) of fluctuations, at the possible price of 
overestimating them if complementing (but overlooked) fluctuation
channels do open up. As a consequence, these simulations are expected 
to be reliable for quadrupole fluctuations or for dynamical  
situations dominated by quadrupole fluctuations, so that a restriction 
to the "collective" fluctuations is reasonable. As shown in \cite{chapelle}
grid simulations in restricted model space indeed validate 
quadrupole fluctuations as calculated in the projection method used here
(see table 1 of \cite{chapelle}). In turn, differences show up 
in the high momentum tail of the distribution function, which are, 
as expected, not necessary accounted for in a multipole-based 
method of reinjection of fluctuations.

Direct kaon and high energy proton production mechanisms are the same in
both ({\bf  $BL_{PM}$ } and QMD) approaches. Whenever two nucleons approach each other closer than
$r = \sqrt{\sigma_{tot} /\pi}$ a NN collision takes place. With a probability
of ${\sigma_{NN \rightarrow X}\over \sigma_{tot}}$
a particle X appears in the exit channel. Thus both, high energy protons
as well as kaons, should reflect the probability to have two body collisions
with very large $\sqrt{s}$ values as compared to the beam energy. Hence
possible differences between the two approaches should
have their origin in a different
momentum distribution of the incoming particles.

\section{Results}

A comparison of the models with experimental results is not that easy
because we are investigating a part of the phase space which is very
dilutely populated. We deal with protons having three times the energy
which is available in a first chance  NN collision
at the same incoming energy per
nucleon and which are emitted in sideward
direction. This requires an extremely high number of simulations of the
reaction.

\begin{figure}[hbt]
%%%
%           who put this dammned waste therein ????????
%%%
%\vspace{-3cm}
%\epsfxsize=16.cm
%$$
%\epsfbox{wirk1n.eps}
%$$
%\vspace{-3cm}
%%%
%           who put this dammned waste herein ????????
%%%
\centerline{\psfig{figure=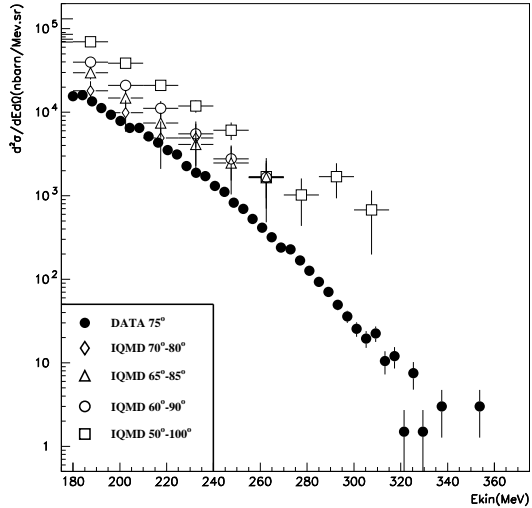,width=0.9\hsize}}

\caption[]{Comparison of the measured proton spectra for Ar (92AMeV)+Ta [9]
to those obtained in QMD using different angular bins.}

\end{figure}

Figure 1 displays
${d\sigma\over d\Omega dE}$ for four different opening angles
in comparison with the experimental results. First of all we
observe that, up to proton kinetic energy of
250 MeV, simulations and experiments agree within error bars
even in this remote phase space region. Despite of 50,000
simulations of the reaction we are not able to simulate directly
the detector acceptance but have to enlarge the acceptance in order
to gain statistics: the diamonds correspond to an acceptance of
$70^\circ \le \theta \le 80^\circ$,
the triangles to one of  $65^\circ \le \theta \le 85^\circ$,
the open circles to one of $60^\circ \le \theta \le 90^\circ$ and
the squares to one of $50^\circ \le \theta \le 100^\circ$.
We see that between $60^\circ$ and $90^\circ$ the cross section
is isotropic in between the error bars  and hence
${d\sigma\over d\Omega dE}$ does not change if we enlarge the opening
angle. If we enlarge the opening angle even further, the slope
stays constant but the absolute value of the cross section increases.
By enlarging the opening angle
the constant slope value should allow for an extrapolation of the
simulated spectra to higher energies 
where QMD calculations do not allow to explore directly high energy protons.

\begin{figure}[hbt]

\centerline{\psfig{figure=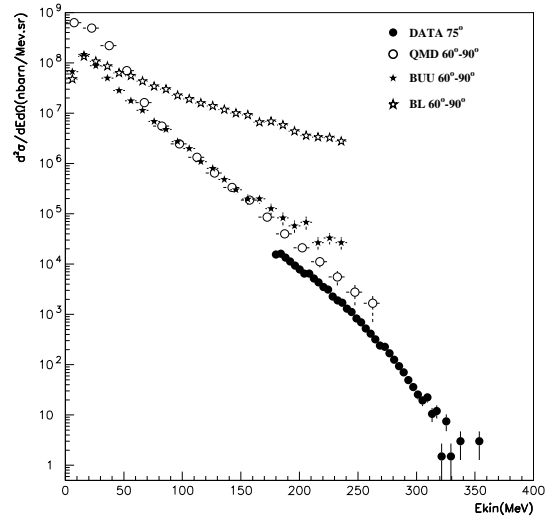,width=0.9\hsize}}

\caption{Proton spectra for Ar (92 AMeV)+Ta  obtained with QMD,
BUU and $BL_{PM}$  in comparison to DATA.}

\end{figure}

Figure 2 compares the double differential cross section
${d\sigma\over d\Omega dE}$ with the results of three different transport
theories, QMD, BUU and {\bf  $BL_{PM}$ }. This figure displays several interesting
features, which need a cautious discussion.  

First of all we observe a quite different slope of the spectra of
the {\bf  $BL_{PM}$ } calculation on the one side and of the BUU and QMD calculations
on the other side. Both slopes are fairly exponential and display
an apparent temperature (in between 200-250 MeV) of 76 MeV and 22 MeV, 
respectively.
The difference of the spectra reaches 3 orders of magnitude at
proton energies of 225 MeV. A comparison with fig 1 shows that this
difference cannot be caused by the enlarged opening angle. We observe
as well, that between $E_{kin}=75$  and  $E_{kin}=200$ MeV, 
BUU and QMD agree, a fact which has
been already discovered a couple of years ago \cite{gyu}.
Below $E_{kin}=50$ MeV BUU and BL coincide and strongly differ
from QMD. This has probably the following reason: For the QMD approach the
spectra of all protons has been displayed (to avoid the rather time
consuming minimum spanning tree algorithm which defines which nucleons
are part of a fragment) whereas in the other two 
approaches the spectra contains only those protons which are finally
not part of a cluster. 

The comparison
to data is also enlightening. First it should be noted that data are
available only in a restricted range of  $E_{kin}$ between about 175 and
350 MeV. Other data \cite{la} from reactions induced by 94 AMeV oxygen 
projectiles show similar proton spectra slopes below 150 MeV. 
 Among the 3 approaches QMD, BUU and {\bf  $BL_{PM}$ }, it is finally QMD which leads, in
a small window between 175 MeV and 250 MeV to the best agreement
with data (apparent temperature of
order 17 MeV) both in absolute values and slope. 

A further point, not directly visible
in the figure, needs also to be mentioned. 
BUU and {\bf  $BL_{PM}$ } give the correct time evolution of the one body distribution
function of the system if one employs an infinite number of test
particles. Therefore the results have been checked on a possible dependence
on the number of employed test particles and we show here numbers 
corresponding to an "asymptotic" regime. It should be noted that 
for small values of the number of test particles we observed a reduced 
yield at high kinetic proton energies.

It is remarkable  that BUU (a one body mean field approach) and QMD 
(a n-body molecular dynamics approach) produce in the interesting region the
same proton spectra and that both agree with experiment, what confirms the
result of  ref. \cite{gyu}. This raises the question whether the
additional fluctuations of {\bf  $BL_{PM}$ } (as compared to BUU) can manifest themselves
in momentum space. Since {\bf  $BL_{PM}$ } introduces fluctuations which are not present
in the BUU mean field approach but contained in QMD by construction
there seems to be little room for this conjecture.
Most probably many body effects
manifest themselves as fluctuations (or correlations) in {\it coordinate
space}. These density fluctuations which are washed out in BUU calculations
are the origin of the clusterization during a heavy ion collision. 
Large differences between QMD and BUU have indeed been
observed \cite{gos} as far as cluster production is concerned.   

\begin{figure}[hbt]
\centerline{\psfig{figure=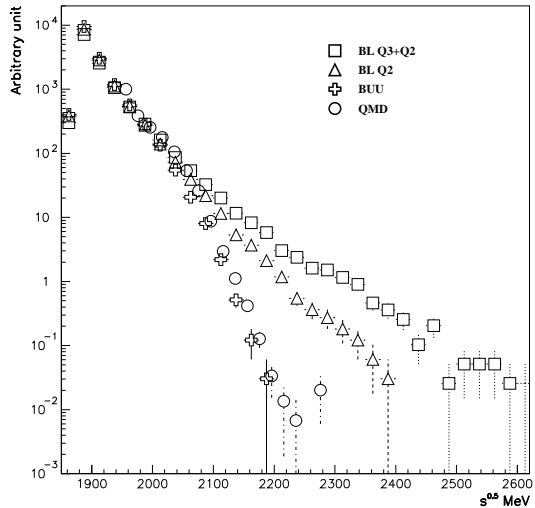,width=0.9\hsize}}

\caption{ %%% \mbox{$\sqrt{s}$} %%% removed due to strange latex-errors
$s^{1/2}$-distribution for Ar(92 AMeV)+Ta at
$b=0$ fm obtained with QMD, BUU and BL.}

\end{figure}

As mentioned above, {\bf  $BL_{PM}$ } overestimates the proton yield at high
kinetic energies.
This strongly enhanced high energy proton component reflects the
larger available energies in the NN center of mass system
in the {\bf  $BL_{PM}$ }  approach, as compared to BUU or QMD.
Figure  3 displays the $\sqrt{s}$ distribution of all NN
collisions performed during the simulation in the various approaches.
We observe a large difference between the models.
In QMD and BUU there is no collision with a $\sqrt{s}$ larger than
2.3 GeV, while in {\bf  $BL_{PM}$ } collisions with an energy beyond the
threshold for kaon production are observed ($\sqrt{s}_{thres} = 2.548$).
The results of {\bf  $BL_{PM}$ } change slightly if one employs
octupole moments (Q2+Q3) as the fluctuating quantities
as compared to the fluctuations of the quadrupole moments (Q2) only.

Independent of the chosen set up we observe, in {\bf  $BL_{PM}$ } simulations,
collisions with much higher $\sqrt{s}$ energies than in the other
simulation programs.
These energetic collisions produce energetic protons with a too large
rate as compared to available data. This probably
happens because the Monte Carlo selection employed to choose the momenta
out of a given
distribution allows that a sizeable fraction of the total momentum is
transferred to a single nucleon. This, as already mentioned, mocks up
"collective" effects.

Subthreshold kaon production is presently a very active research field 
\cite{ha,ko,gi} because kaons are considered as a possible messenger
from the high density zone of the reaction and hence of a possible onset
of the chiral phase transition. Up to recently, however, it was not believed
that below 600 AMeV kaons could be observed. Experimentally as well as
theoretically the probability to find a kaon becomes exponentially low with
decreasing energy and hence the beam or calculation time exceeds present
possibilities. 

 A while ago it has been
reported that kaons have been observed in a heavy ion
reaction at an energy as low as 92 AMeV \cite {ju}. Recently this experiment
has been repeated with a comparable result ($\sigma_{K^+} =2.9 \pm 1.6 
\cdot 10^{-9}b$
for the reaction Ar + Ta at 92 MeV/N) \cite{le}.
It should be noted that the necessary center of mass energy to produce
a kaon is 671 MeV + twice the mass of the nucleons.
Hence, in the investigated system,$^{36} Ar + ^{48} Ti$ at 92 AMeV, which has
a $\sqrt{s}-84 m_p$ of 2.0 GeV, 34\% of the total available energy is
needed to create a kaon. This points towards a highly collective process.
We see that {\bf  $BL_{PM}$ }, in the set up Q2+Q3, reproduces these data. We find
4 pp collision above threshold. In these collisions we produce a kaon
with a  probability of about $0.25 \cdot 10^{-3}$\cite{cosy}. This number
has to be divided by $2 \cdot 10^{5}$ for the number of test particles
and the number of events and has to be multiplied by 1.3 b for the
total reaction cross section. This yields a kaon production cross
section of $1.6 \cdot 10^{-9}b$ which is comparable with the experimental
result. This argument should nevertheless be taken with some due caution 
in view of the very small number of relevant pp collisions. Still 
it qualitatively provides a coherent picture with the results of \cite{be}.
    
It should be noted that the estimates of \cite{be} rely on a schematic 
model, built from BL simulations based on quadrupole fluctuations. 
Only characteristics of the quadrupole do serve as inputs of the model, 
which can thus be considered as numerically safe (see the above discussions
on \cite{chapelle} and internal checks, for example with  respect to BUU, 
as presented in \cite{be}). The model of \cite{be} is thus, by 
construction, highly collective: it presumably exhausts a sizeable 
part of the collective source of kaon yield. But it is of course by no means 
a direct simulation of the BL equation. In this respect, kaon production 
in QMD, together with the collective schematic model of \cite{be}, 
provide a coherent picture, namely the fact that a dominant fraction 
of kaon yield does stem from collective effects. In turn, high energy proton 
production seems to originate from direct (non collective) incoherent 
two-body processes. 

From the above calculations, we thus see that models which
describe the proton spectra quantitatively have problems with describing
kaon production based on binary collisions. Conversely, models
accommodating collective effects, raise problems within explaining
high energy proton spectra. This may point to a collective
kaon production mechanism not included in the QMD or BUU approach
or to strong in medium modification of the properties of strange
particles (like a lowering of the  mass) which may lower the threshold
for the production. Calculations predict, however, that in nuclear matter
the kaon production threshold increases \cite{scha}. Therefore it
is rather unlikely that modifications of the elementary production process
can be the reason for the very subthreshold kaon production.
One could argue that collective effects 
should also affect the pion production, which is observed \cite{cas90} to originate
from the participant zone. There, however, it is found that the number of 
observed pion scales well with the number of nn collisions above threshold
\cite{cas90}.

\section{Conclusions}
In summary, we can conclude that high energy proton spectra are reproduced
by QMD calculations, and to a lesser extent by BUU. Both are not able
to reproduce well below threshold kaon production. In turn,  {\bf  $BL_{PM}$ }
fails to explain high energy proton spectra but either direct
\cite{suraud} or schematic \cite{be} fluctuations reproduce the order
of magnitude of the kaon production. 
{\bf  Whether this is only a consequence the realization of the BL
equations we use or an inherent problem of the BL approach can only be judged if a
different realization is developed. The fact that QMD and BUU have already 
sufficient fluctuations in momentum space may lead to the conjecture that 
the Langevin force should generate fluctuations in coordinate space only.
How this is possible remains to be seen.} 
We can conclude that high energy protons do not
call for highly collective effects, contrarily to well below threshold kaons.

%%\vspace*{0.2cm}\\

%%%%%%%%%%%%%%%%%%%%%%%% newpage %%%%%%%%%%%%%%%%%%%%%%%%%%%%%
%\pagebreak


\begin{references}

\bibitem{tdhf} K.T.R Davis et al., in Heavy Ion Science, Ed. D.A. Bromley,
1984, Plenum Press NY

\bibitem{ai85a}
J.~Aichelin and G.~Bertsch.
Phys.~Rev.~{\bf C31}, 1730 (1985).

\bibitem{cas90}
W.~Cassing, V.~Metag, U.~Mosel and K.~Niita.
Phys.~Reports~{\bf 188}, 361 (1990).

\bibitem{st86}
H.~St\"ocker and W.~Greiner. Phys.~Reports~{\bf 137}, 277 (1986).

\bibitem{bonas}
A.~Bonasera et al., Phys. ~Reports~{\bf 243}, 1 (1994)

\bibitem{greg87}
C.~Gregoire, B.~Remaud, F.~Sebille, L.~Vinet, and Y.~Raffray,
Nucl. Phys. {\bf A465}, 317 (1987).


\bibitem{su} Y. Abe et al., Phys. Rep.{\bf 275} (1996) 49, and references 
therein

\bibitem{gr} S. Ayik Z. Phys. {\bf A298} (1980) 83 and S. Ayik et al.
Nucl. Phys. {\bf A 513} (1990) 187
\bibitem{ra}J. Randrup et al., Nucl. Phys. {\bf A 514} (1990) 339
\bibitem{re} G.F. Burgio et al. Nucl. Phys. {\bf A 529} (1991) 157 
\bibitem{be} M Belkacem et al, Phys. Rev. {\bf C47}, (1993) R 16
\bibitem{ge} M. Germain et al.,   Nucl. Phys. {\bf A620} (   1997) 80
\bibitem{ai} J. Aichelin, Phys. Rep. {\bf 202}, 233 (1991).
\bibitem{ha} C. Hartnack et al., Nucl. Phys. {\bf A580} (1994) 643
\bibitem{iqmd} Ch. Hartnack, GSI-report 93-05, S.A. Bass et al. Phys. Rev.
{\bf C51} (1995) 3343, Ch. Hartnack et al. submitted to Z. Phys. A
\bibitem{suraud} E. Suraud et al, Nucl. Phys. {\bf A542} (1992) 141
\bibitem{chapelle} F. Chapelle et al, Nucl. Phys. {\bf A540}(1992)227
\bibitem{gyu} J. Aichelin et al., Phys. Rev. Lett. {\bf 62} (1989) 1461
\bibitem{la} J.L. Laville et al., Nucl. Phys {\bf A564} (1993) 564
\bibitem{gos} P.B. Gossiaux et al., Phys. Rev. {\bf C51} (1995) 3357
\bibitem{ko}G.Q. Li et al., Phys. Rev {\bf C57} (1998) 434
\bibitem{gi}E.L. Bratkovskaya et al. Nucl. Phys. {\bf A622} (1997) 593
\bibitem{ju} J. Julien et al., Phys. Lett. {\bf B264}, 269 (1991)
\bibitem{le} F.R. Lecolley, Nucl. Phys {\bf A 583}, 379c (1995)
\bibitem{cosy} S. Brausiepe et al., Preprint KfA J\"uelich 
(m.wolke $@$ kfa-juelich.de)
\bibitem{scha} J. Schaffner-Bielich et al. Nucl. Phys. {\bf A625}
(1997) 325

\end{references}
\end{document}